\documentclass{aa}
\usepackage{amsmath}
\usepackage[varg]{txfonts}

\usepackage{natbib}
\usepackage{graphicx}
\usepackage{dcolumn}
\usepackage{aas_macros}

\newcommand{\reaction}[6]{\nuc{#1}{#2}(#3,#4)\/\nuc{#5}{#6}}
\newcommand{\nuc}[2]{\ensuremath{^{#1}}#2}
\newcommand{\Erlab}[1]{\ensuremath{E_{r}^{\textrm{lab}} = #1}~keV}
\newcommand{\Ercm}[1]{\ensuremath{E_{r}^{\textrm{c.m.}} = #1}~keV}

\graphicspath{{./figs/}}

\begin{document}


\title{Correlated Uncertainties in Monte Carlo Reaction Rate
  Calculations}

\author{Richard Longland\inst{1,2}}
\institute{North Carolina State University, Raleigh, NC 27695 
  \and  Triangle Universities Nuclear Laboratory, Durham, NC 27708
}

\date{\today}

\abstract{Monte Carlo methods have enabled nuclear reaction rates from
  uncertain inputs to be presented in a statistically meaningful
  manner. However, these uncertainties are currently computed assuming
  no correlations between the physical quantities that enter those
  calculations. This is not always an appropriate
  assumption. Astrophysically important reactions are often dominated
  by resonances, whose properties are normalized to a well-known
  reference resonance. This insight provides a basis from which to
  develop a flexible framework for including correlations in Monte
  Carlo reaction rate calculations.}  {The aim of this work is to
  develop and test a method for including correlations in Monte Carlo
  reaction rate calculations when the input has been normalized to a
  common reference.}{A mathematical framework is developed for
  including correlations between input parameters in Monte Carlo
  reaction rate calculations. The magnitude of those correlations is
  calculated from the uncertainties typically reported in experimental
  papers, where full correlation information is not available. The
  method is applied to four illustrative examples: a fictional
  3-resonance reaction, \reaction{27}{Al}{p}{$\gamma$}{28}{Si},
  \reaction{23}{Na}{p}{$\alpha$}{20}{Ne}, and
  \reaction{23}{Na}{$\alpha$}{p}{26}{Mg}.}{Reaction rates at low
  temperatures that are dominated by a few isolated resonances are
  found to minimally impacted by correlation effects. However,
  reaction rates determined from many overlapping resonances can be
  significantly affected. Uncertainties in the
  \reaction{23}{Na}{$\alpha$}{p}{26}{Mg} reaction, for example,
  increase by up to a factor of 5. This highlights the need to take
  correlation effects into account in reaction rate calculations, and
  provide insight into which cases are expected to be most affected by
  them. The impact of correlation effects on nucleosynthesis is also
  investigated.}{}

\keywords{methods: numerical -- methods: statistical -- nuclear reactions, nucleosynthesis, abundances}
\maketitle

\section{\label{sec:intro}Introduction}

Observing stellar phenomena provides insight into the evolution of
matter in the universe. However, these observations are necessarily
constrained by the opacity of stellar atmospheres to radiation; we
cannot easily probe deep into the cores of stars to directly observe
their structure. Stellar models are required to make the connection
between observations and the internal structure of stars. Clearly, the
accuracy of these models is paramount. To fully investigate the
validity of stellar models and compare them with observations, we must
account for uncertainties in the physical quantities used as inputs to
those models. One such input is the rate at which nuclear reactions
occur in stellar environments. Understanding the uncertainties in
reaction rates is therefore critical to understanding the internal
structure of stars and stellar phenomena.

A large step forward in determining statistically realistic
uncertainties of reaction rates was made in
\cite{LON10,ILI10b,ILI10c}, and \cite{ILI10a}. A Monte Carlo method
was developed for determining statistically realistic uncertainties of
reaction rates taking into account experimental uncertainties. This
was then expanded by \cite{SAL13} and \cite{MOH14}, and summarized in
\cite{ILI15} and \cite{CHA14}. These developments form the basis for
the Starlib reaction rate library (starlib.physics.unc.edu). Similar
methods have been developed by \cite{RAU16}. Briefly, nuclear reaction
cross section inputs are represented by probability density
distributions, which are then propagated, through Monte Carlo
variation, to a probability density distribution of the reaction
rate. This final reaction rate distribution is temperature
dependent. Unlike more traditional methods that yield ``upper'' and
``lower'' limits on reaction rates, this method yields continuous
probability functions with a ``high'' and ``low'' rate defined by the
68\% coverage interval. Once uncertainties of reaction rates are
determined, they can be used in stellar models to determine the effect
that nuclear physics uncertainties have on nucleosynthesis. It should
be stressed here that these uncertainties can be either derived from
experimental results or from estimated theoretical uncertainties.

Often, cross sections are dominated by resonances. Hence, the
uncertain cross section inputs in reaction rate calculations are
resonance strengths, partial width measurements, or resonance
energies. These are often not measured absolutely, but are measured
with respect to well-known ``reference'' resonances. Indeed, the first
reaction rate compilation based on reference resonances was that of
\cite{ILI01}. These reference resonances often have appreciable
uncertainties of their own, leading to correlated uncertainties in the
relative measurements. The nature of the correlation between
quantities is often not reported. However, a conservative estimate of
this effect can be estimated by comparing the magnitude of measurement
uncertainties. For example, if a standard resonance strength is
uncertain by 20\%, then all subsequent measurements are also uncertain
by 20\% \textit{plus} any additional systematic or statistical
uncertainty present in their measurement. Those resonance strengths
are correlated. Accounting for this effect has, so far, not been
included in the Monte Carlo reaction rate formalism discussed
above. Here, an extension to that method is described that allows
correlated uncertainties to be taken into account in reaction rate
calculations.

This paper is organized as follows: an introduction to the Monte Carlo
reaction rate formalism is presented in Sect.~\ref{sec:monte-carlo}
before the addition of correlated uncertainties is discussed in
Sect.~\ref{sec:correlations}. The correlation formalism for Monte
Carlo reaction rates is developed in
Sect.~\ref{sec:correlations-montecarlo}. Four test cases are
used to investigate the impact of these effects on nucleosynthesis
models. Their results are discussed in
Sect.~\ref{sec:results-discussion}, and conclusions are presented in
Sect.~\ref{sec:summary}.

\section{Monte Carlo Reaction Rates}
\label{sec:monte-carlo}

The reaction rate per particle pair, $\langle \sigma v \rangle$, is
defined as
\begin{equation}
  \label{eq:reactionrate}
  \langle \sigma v \rangle = \left(\frac{8}{\pi \mu}\right)^{1/2}
  \frac{1}{(k_B T)^{3/2}}\int_0^{\infty} E \sigma(E) e^{-E/k_BT} dE,
\end{equation}
where $\mu$ is the reduced mass of the system, $k_B$ is the
  Boltzmann constant, $T$ is the temperature at which the reaction
rate is being calculated, and $\sigma(E)$ is the energy-dependent
cross section of the reaction. Many reactions of astrophysical
importance proceed through nuclear resonances, populating compound
nuclear states that subsequently decay. For a single, isolated
resonance, the cross section in Eq.~(\ref{eq:reactionrate}) can be
replaced by
\begin{equation}
  \label{eq:resonance-xsec}
  \sigma(E) = \frac{2 J + 1}{(2 J_1 + 1)(2 J_2 + 1)} \frac{\pi}{k_B^2}\frac{\Gamma_a(E) \Gamma_b(E)}{(E-E_r)^2 + \Gamma^2/4}.
\end{equation}
Here, $J$, $J_1$, and $J_2$ are the spins of the resonance, target,
and projectile particles, respectively.  $\Gamma_a$ and $\Gamma_b$ are
energy-dependent quantities describing the particle and reaction
partial widths of the state in question. For example, for a
(p,$\gamma$) reaction, $\Gamma_a$ corresponds to the proton partial
width and $\Gamma_b$ is the $\gamma$-ray partial width. $\Gamma$
corresponds to the \textit{total} width of the state, and $E_r$ is the
resonance energy. These parameters are often determined experimentally
using a wide variety of experimental techniques of differing
precision. Some experimental methods are best suited to measuring the
integrated cross section across narrow resonances. In that case, we
refer to a resonance strength, $\omega \gamma$, and
Eq.~(\ref{eq:reactionrate}) is replaced with
\begin{equation}
  \label{eq:reactionrate-narrowresonance}
  \langle \sigma v \rangle = \left(\frac{2\pi}{\mu k_BT}\right)^{3/2}
  \hbar^2 \sum_i \omega \gamma_i e^{-E_r/k_BT}
\end{equation}
This paper focuses on reaction rates dominated by resonances in the
absence of interference. Addressing interfering resonance reaction
rates that are best described by R-matrix or other complex models is
left for future work.

Each of the partial widths, resonance strengths, and resonance
energies described by Eqns.~(\ref{eq:resonance-xsec}) and
(\ref{eq:reactionrate-narrowresonance}) has some associated
uncertainty. These are used as inputs to calculate reaction rate. The
general strategy for using Monte Carlo uncertainty propagation of
reaction rates is as follows: (i) generate random variables
corresponding to each uncertain input by varying them according to
their experimental uncertainties, (ii) calculate the reaction rate at
each temperature, and (iii) repeat many (10,000) times. Care must be
taken during this procedure to correctly propagate resonance energy
uncertainties, particularly for wide resonances that are integrated
numerically. The resonance energy enters Eq.~(\ref{eq:resonance-xsec})
in multiple places, so this must be taken into account. Following this
procedure, an ensemble of reaction rates is obtained that can be
summarized using descriptive statistics. \cite{LON10} found that two
parameters are sufficient to summarize the reaction rate at each
temperature, that is, $\mu$ and $\sigma$, the location and shape
parameters of a lognormal distribution. Note that these parameters
evolve smoothly as a function of temperature. The recommended reaction
rate at temperature $T$ is given by (see \cite{EVA00,SAL13})
\begin{equation}
  \label{eq:recommendedrate}
  \langle \sigma v \rangle_{\textrm{rec.}} = e^{\mu(T)},
\end{equation}
and the factor uncertainty, $f.u.$ is defined by
\begin{equation}
  f.u. = e^{\sigma(T)}.
\end{equation}
Additionally, the ``low'' and ``high'' rates provided by the
1-$\sigma$ uncertainties are found using
\begin{equation}
  \label{eq:highlowrate}
  \langle \sigma v \rangle_l = \langle \sigma v \rangle_{\textrm{rec.}}
  \, (f.u.)^{-1} \qquad \langle \sigma v \rangle_h = \langle \sigma v
  \rangle_{\textrm{rec.}} \, (f.u.)^{+1}
\end{equation}

Once parameters that describe temperature dependent reaction rates and
their uncertainties have been found, Monte Carlo procedures can also
be applied to find the effect these uncertainties have on
nucleosynthesis in stars. Techniques for using the
temperature-dependent recommended rate and factor uncertainty
parameters in nucleosynthesis calculations were first investigated in
\cite{LON12c}. A single, randomly sampled temperature-dependent
parameter, $p(T)$, was found to be sufficient for generating reaction
rate samples. In fact, it was also found that $p(T)$ need not be
temperature dependent for accurate reproduction of nucleosynthesis
uncertainties, and must only be normally distributed. Thus, a
  single reaction rate sample can be represented by
\begin{equation}
  \label{eq:ratesample-simple}
  \langle \sigma v \rangle_i = \langle \sigma v \rangle_{\textrm{rec.}}
  \, (f.u.)^{p_i}.
\end{equation}

Monte Carlo nucleosynthesis calculations can be performed using this
scheme by generating an ensemble of standard normally distributed
$p_i$ values for each reaction. As long as those values are retained
for each nucleosynthesis calculation, techniques can be developed to
characterize which reaction rate uncertainties most affect model
predictions of nucleosynthesis. Most of these methods consist of
calculating the correlation between the final abundance of a
particular isotope with the $p_i$ values of a reaction rate. A few of
those techniques have already been investigated by \cite{ILI15}, and
further methods are forthcoming. It should be stressed, here, that
these $p_i$ values represent variations of a reaction
\textit{within its experimental or theoretical uncertainty}.
Sect.~\ref{sec:results-discussion} relies on these Monte Carlo
nucleosynthesis techniques to investigate the impact of correlated
nuclear inputs on nucleosyhtnesis predictions in stellar modes. For
the sake of clarity, only the reaction in question and its reverse
reaction are varied, holding all other reactions at their recommended
rates.

As mentioned, the first step in the procedure to calculate Monte Carlo
reaction rate uncertainties is to generate random variables for all
input parameters. This step provides us with an opportunity to account
for input parameter correlations. Procedures already exist for
generating correlated random numbers and are laid out in
Sect.~\ref{sec:correlations-montecarlo}, but first the source and
behavior of those correlations should be determined.

\section{Correlated Uncertainties in Reaction Rates}
\label{sec:correlations}



Resonance strengths can be measured directly. A solid substrate
containing the target of interest is usually constructed, and
bombarded by a beam of particles corresponding to the incoming
channel. Note that the concepts outlined in this section are equally
applicable to inverse kinematics experiments. The reaction yield is
then measured using some detection scheme. For example, a high
intensity proton beam can be used in conjunction with $\gamma$-ray
detectors to measure proton radiative capture reactions. To determine
a single resonance strength, the beam energy is tuned such that the
beam particles lose energy as they traverse the target, thus
effectively integrating the cross section (a more in-depth discussion
of these techniques can be found in
\cite{CAULDRONS} and \cite{ILIBook}). The resonance strength,
$\omega \gamma$, is determined by
\begin{equation}
  \label{eq:narrowyield}
  \omega \gamma = \frac{2 \varepsilon_r}{\lambda_r^2} \frac{N}{N_b B
    \eta W},
\end{equation}
where $\varepsilon_r$ is the stopping power of the target (usually
replaced by $\varepsilon_{\textrm{eff}}$ if the target is a compound
substrate) and $\lambda_r$ is the deBroglie wavelength at the
resonance energy. The quantities $N$ and $N_b$ denote the number of
detected products ($\gamma$-rays in this case) and number of beam
particles (protons), respectively. $B$ is the branching ratio of the
detected reaction product, $\eta$ is the detection efficiency, and $W$
is a quantity that takes into account any angular-dependent effects in
the reaction. Every one of these parameters is uncertain to some
degree. The uncertainty in a resonance strength is therefore computed
from the product of random variables. The Central Limit Theorem tells
us that the uncertainty in this product will be distributed according
to a lognormal distribution. This concept and its impact on
nuclear reaction rate uncertainties is described in more detail by
\cite{LON10}.

To determine a resonance strength absolutely, target material
properties must be known to a high degree of accuracy. 
Beam current and absolute detector efficiencies must be similarly well
known. To avoid these constraints, however, one can turn to relative
measurements. Here, a resonance strength measurement is normalized to
a well known resonance that was measured, separately, using a
carefully calibrated system. That resonance - referred to here as a
\textit{reference resonance} and denoted herein with a subscript `r' -
is then used to normalize subsequent measurements. Absolute
determination of beam current, detector efficiency, and target
properties can then be canceled in Eqn.~(\ref{eq:narrowyield}). The
consequence of this procedure is that those subsequent resonance
strengths are correlated with the reference resonance. Furthermore,
their uncertainties are necessarily \textit{larger} than the reference
strength uncertainty. This fact is crucial to the correlation
technique described below. For example, \cite{POW98} measured the
strength of the \Erlab{406} resonance in
\reaction{27}{Al}{p}{$\gamma$}{28}{Si} with an uncertainty of
$\sigma_r = 6.0$\%. Any subsequent relative resonance strength
measurement in the same reaction will contain an uncertainty
\begin{equation}
  \label{eq:uncertainty-propagation}
  \sigma^2 = \sigma_j^2 + \sigma_r^2,
\end{equation}
where $\sigma_j$ comprises of any uncertainty in the relative
measurement that cannot be cancelled out in
Eqn.~\ref{eq:narrowyield}. Any remaining factors affecting the
uncertainty in resonance $j$ will, therefore, be \textit{independent}
of resonance $r$. Equation~(\ref{eq:uncertainty-propagation}) dictates
that all normalized resonances must have larger uncertainties than the
reference. This also gives us a convenient method for identifying the
reference resonance when such information is not clearly presented:
the reference resonance will always contain the smallest
uncertainties.

Similar arguments can be made for partial width determinations. Often,
if a partial width is determined based on a spectroscopic factor or
Asymptotic Normalization Coefficient (ANC), normalization to a well
known reference resonance is performed. In the interest of clarity,
the following discussion will only refer to reaction rates that are
dominated by narrow resonance strengths.

\subsection{Monte Carlo Treatment of Correlated Uncertainties}
\label{sec:correlations-montecarlo}

To demonstrate the generation of correlated Monte Carlo uncertainties
in resonance strengths for reaction rate calculations, we first
consider correlated random variables represented by the vector,
$\mathbf{x'}$. Assuming, for now, that $\mathbf{x'}$ is distributed
according to a multivariate probability density function characterized
by mean values of $\boldsymbol{\mu}$ and standard deviations of
$\boldsymbol{\sigma}$, we can write
\begin{equation}
  \label{eq:multivariate}
  P_{\mathbf{x'}} = \frac{1}{\sqrt{(2\pi)^k |\boldsymbol{\Sigma}|}}
  \exp\left(-\frac{1}{2}(\mathbf{x'}-\boldsymbol{\mu})^T \boldsymbol{\Sigma}^{-1}
    (\mathbf{x'}-\boldsymbol{\mu}) \right).
\end{equation}
The matrix, $\boldsymbol{\Sigma}$ is the covariance matrix, which is
assumed to be positive definite and real.  Typically, a reaction rate
is calculated from the contributions of a large ensemble of
resonances. The covariance matrix can therefore be made up of a
complex interplay of uncertainties and experimental effects. However,
this picture is simplified by realizing that narrow resonance
strengths are almost always normalized to a single reference
resonance, as already discussed. In this case, the covariance matrix
simplifies considerably with no cross correlations. Thus, the
uncertainty of each resonance strength is represented by an individual
bivariate correlation, but with a different correlation coefficient.

For the bivariate case, $\mathbf{x'}=[x,y]$ and the covariance matrix
is defined by
\begin{equation}
  \label{eq:bivariate-covarience}
  \boldsymbol{\Sigma} =
  \begin{bmatrix}
    \sigma_x^2 & \rho \sigma_x \sigma_y \\
    \rho \sigma_y \sigma_x & \sigma_y^2 
  \end{bmatrix}
  = \mathbf{L L^T},
\end{equation}
where $\rho$ is the correlation coefficient, which varies between 0
and 1; $\mathbf{L}$ is a lower triangular matrix with real, positive
diagonal entries; and $\mathbf{L^T}$ is its conjugate transpose. This
decomposition is the so-called \textit{Cholesky decomposition}, and
$\mathbf{L}$ can be computed using \citep{STE00}
\begin{align}
  \label{eq:cholesky-decomposition}
  L_{jj} =& \sqrt{\Sigma_{jj} - \sum_{k=1}^{j-1}L_{jk}^2}, \\
  L_{ij} =& \frac{1}{L_{jj}} \left(\Sigma_{ij} -
            \sum_{k=1}^{j-1}L_{ik}L_{jk} \right), \quad \textrm{for } i>j.
\end{align}
For Eq.~(\ref{eq:bivariate-covarience}), $\mathbf{L}$ becomes
\begin{equation}
  \label{eq:bivariate-L}
  \mathbf{L} =
  \begin{bmatrix}
    \sigma_x & 0 \\
    \rho \sigma_x  & \sigma_y \sqrt{1-\rho^2}.
  \end{bmatrix}
\end{equation}

This Cholesky decomposition is useful because it can be used in Monte
Carlo calculations to convert uncorrelated normally distributed
random variables, $\mathbf{x}$, into correlated quantities, $\mathbf{x'}$,
by \citep{REU11}
\begin{equation}
  \label{eq:correlating-uncorrelated}
  \mathbf{x'} = L \mathbf{x}.
\end{equation}

Equations (\ref{eq:correlating-uncorrelated}) and
(\ref{eq:bivariate-L}) can be applied to our bivariate case. Here, we
make the additional simplifying assumption that our variables are
standardized. That is, $\sigma_x = \sigma_y = 1$. In the discussion
above, the concept of reference resonances was discussed. Here, we
introduce that concept into our mathematical description of correlated
Monte Carlo reaction rates. In the following, the parameter $x$
represents the reference resonance, and $y$ is the other resonance in
question that has been normalized to that reference. For sample $i$ of
resonance $j$, therefore, we obtain
\begin{equation}
  y_{j,i}' = \rho_j x_{i} + \sqrt{1-\rho_j^2}\, \, y_{j,i}. \label{eq:yp}
\end{equation}

Why assume $\sigma = 1$? Often in statistical procedures, particularly
those involving Monte Carlo computation, it is convenient to
standardize parameters to avoid scale effects. It is trivial to
un-standardize the random variables following their computation in
Eq.~(\ref{eq:yp}). For example, since resonance strengths are random
variables described by a \textit{lognormal} probability density
distribution as discussed above, Monte Carlo samples of a resonance
strength, $\omega \gamma_j$, can be described in the same way
  as the reaction rate in Eq.~(\ref{eq:ratesample-simple}) by
\begin{equation}
  \label{eq:res-strength-sample}
  \omega \gamma_{j,i} = \omega \gamma_{j,\textrm{rec.}} \, (f.u.)^{y'_{j,i}}
\end{equation}
where $\omega \gamma_{j,\textrm{rec.}}$ is the recommended
  resonance strength, and $f.u.$ is its factor uncertainty. The
quantity $y'_{j,i}$ is a standard, normally distributed, correlated
random variable calculated in Eqn.~(\ref{eq:yp}). To fully describe
correlated resonance strengths, therefore, one only needs to compute
the values $y'_{j,i}$ using Eq.~(\ref{eq:yp}).

\begin{figure*}
  \centering
  \includegraphics[width=0.7\textwidth]{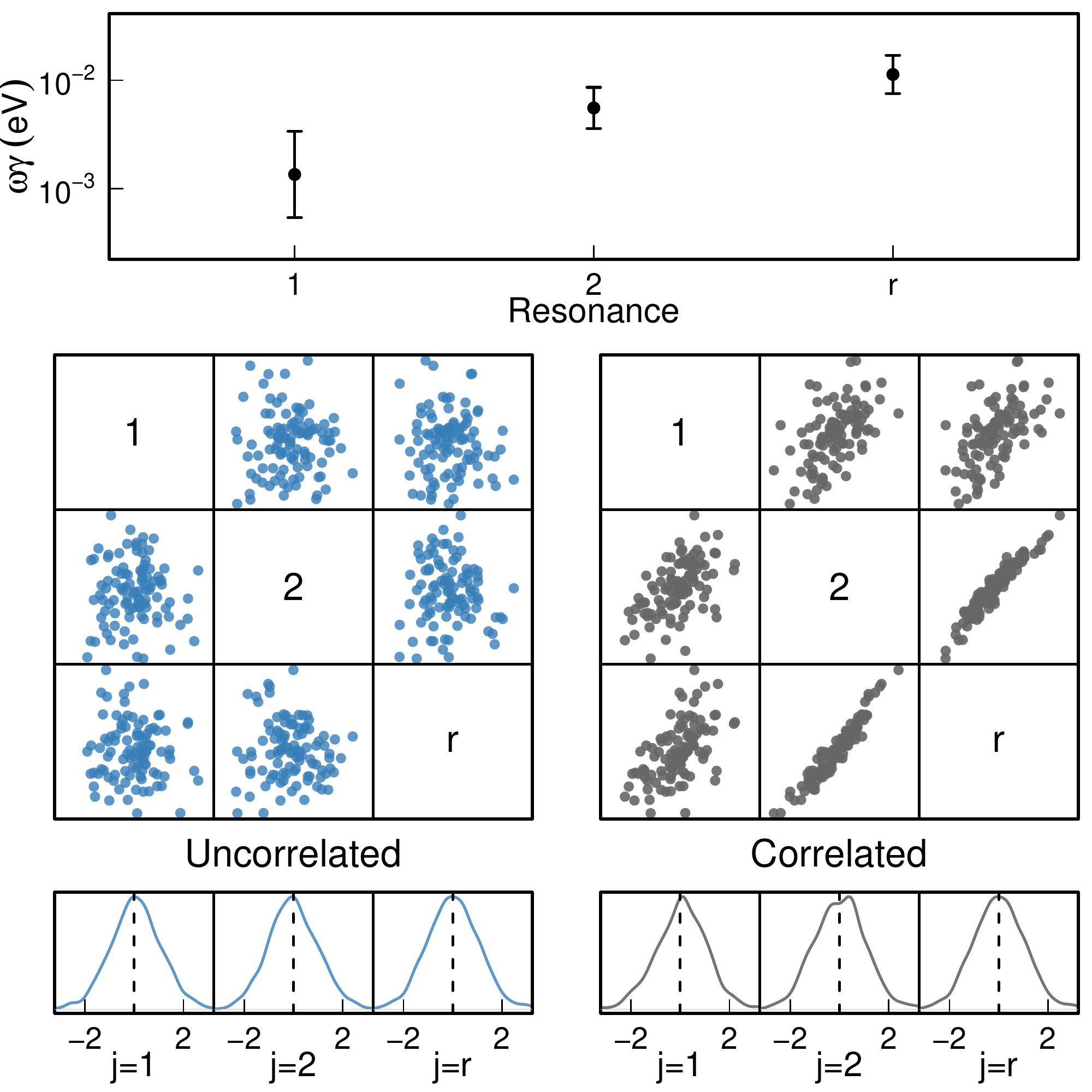}
  \caption{(color online) Example of correlated Monte Carlo
    samples. The top panel shows the strengths of our three example
    resonances including their uncertainties. The middle panels show
    the uncorrelated (blue - $y_j$) and correlated (grey - $y'_j$)
    normally distributed samples, which are used to compute the
    resonance strengths in Eq.~(\ref{eq:res-strength-sample}). Note
    that after applying the correlation outlined in the text,
    resonances 2 and $r$ become highly correlated whereas resonance
    one remains largely unchanged. The lower two panels display the
    density distributions of the samples, and confirms that the
    individual distributions do not vary outside of statistical
    fluctuations.}
  \label{fig:CorrelationDemo}
\end{figure*}


The remaining challenge is to determine the correlation
parameters, $\rho_j$, for each resonance. In the absence of
information, it would be pertinent to assume, conservatively, that
resonances are maximally correlated. We cannot, however, simply use
$\rho=1$. Recall that any normalized quantity must have larger
uncertainties than the reference resonance. Therefore, the correlation
parameter for each resonance strength is calculated by the ratio of
their fractional uncertainties:
\begin{equation}
  \label{eq:rho}
  \rho_j = \frac{\sigma_r}{\omega
    \gamma_r} \frac{\omega \gamma_j}{\sigma_j} = \frac{f.u._r}{f.u._j}.
\end{equation}
Note that the fractional uncertainties can be simply
  replaced by the factor uncertainty.  When the resonance,
$j$, contains an uncertainty close to the reference, $r$, its
uncertainty is dominated by the uncertainty in the latter and
Eq.~(\ref{eq:rho}) shows that $\rho_j \rightarrow 1$. Furthermore, as
statistical and other uncorrelated uncertainties begin to dominate
resonance strengths, their uncertainties become much larger than those
of the reference strength and $\rho$ becomes small. Under this set of
assumptions, it's also impossible for $\rho$ to exceed unity.

The general procedure for calculating correlated reaction rate
uncertainties from narrow resonance strengths is therefore as follows:
(i) identify the reference resonance, $r$ by assuming it is the
resonance with the smallest fractional uncertainty; (ii) calculate the
correlation parameters for all other resonance strengths using
Eq.~(\ref{eq:rho}); (iii) generate a set of standard normally
distributed random variables, $y_{j,i}$ for each resonance; (iv)
correlate those using Eq.~(\ref{eq:yp}) to obtain $y'_{j,i}$; (v)
calculate an ensemble of correlated resonance strengths using
Eq.~(\ref{eq:res-strength-sample}); and (vi) compute the reaction rate
for each set of samples. The ensemble of reaction rates from this
procedure can be summarized to find a recommended rate, ``High'' and
``Low'' rates, and shape parameters $\mu$ and $\sigma$ using the
procedures discussed in \cite{LON10}.

\section{Results and Discussion}
\label{sec:results-discussion}


To investigate the effects of correlated cross section uncertainties
on reaction rates, a few cases are considered. A simple example
reaction consisting of three resonances will be used to illustrate the
procedure, followed by \reaction{27}{Al}{p}{$\gamma$}{28}{Si} and two
examples of reactions on \nuc{23}{Na}:
\reaction{23}{Na}{p}{$\alpha$}{20}{Ne} and
\reaction{23}{Na}{$\alpha$}{p}{26}{Mg}.

\subsection{Simple Example}

To illustrate the procedure of generating correlated uncertainties in
Monte Carlo reaction rates, consider a fictional reaction consisting
of three resonances with strengths $\omega \gamma_1$,
$\omega \gamma_2$, and the reference resonance, $\omega \gamma_r$. The
strengths of the first two resonances are normalized to the reference
resonance and we assume that their uncertainties have been correctly
computed to account for that. Details of their energies and resonance
strengths can be found in table~\ref{tab:fictionalreaction}, and are
plotted in the top panel of Fig.~\ref{fig:CorrelationDemo}. The
resonance strengths all have different uncertainties owing to expected
experimental constraints. The weaker resonance, for example, has a
large uncertainty arising mostly from counting
statistics. 
Inspection of the uncertainties, therefore, immediately reveals that
resonance 2 is highly correlated with the reference resonance, whereas
resonance 1 is dominated by low statistics, and is therefore less
correlated. The correlation factors calculated from
Eqn.~(\ref{eq:rho}) reflect these arguments.

\begin{table}[h]
  \centering
  \begin{tabular}{c|cccc}
    \hline \hline
    Index & E$_r^{c.m.}$ & $\omega \gamma $      & Factor      & Correlation \\
          & (keV)        & (eV)                  & Uncertainty & $\rho$      \\ \hline
    1     & 150          & $1.35 \times 10^{-3}$ & 2.5         & 0.60        \\
    2     & 240          & $5.55 \times 10^{-3}$ & 1.55        & 0.98        \\
    $r$   & 260          & $1.13 \times 10^{-2}$ & 1.50        & 1.00        \\ 
    \hline \hline
  \end{tabular}
  \caption{Resonance parameters for example reaction used to
    demonstrate the effect of correlated uncertainties in resonance
    strengths on a reaction rate uncertainties. Resonance energy
    uncertainties are assumed to be small in this example. Factor
    uncertainties in the resonance strengths are given in the 4th
    column, and the resulting correlation coefficients, $\rho$, are
    shown in the 5th column.}
  \label{tab:fictionalreaction}
\end{table}

The procedure outlined in Sec.~\ref{sec:correlations-montecarlo} is
performed, and shown in the middle row of
Fig.~\ref{fig:CorrelationDemo}. The \textit{uncorrelated} values,
$y_{j,i}$ are generated and shown in the left-middle panel in a
``pairs plot'': a scatterplot matrix showing the bivariate
relationships between the resonance strengths \citep{HAR75}.  The
\textit{correlated} values calculated from Eq.~(\ref{eq:yp}) using the
correlation parameters listed in table~\ref{tab:fictionalreaction} are
displayed in the right-side pairs plot. It is immediately apparent
that the Monte Carlo samples for resonance 2 become highly correlated
with the reference resonance, but there is only weak correlation for
resonance 1. Note, also, that the probability density distributions
for individual parameters remain unchanged following the correlation
procedure, as shown in the bottom row of
Fig.~\ref{fig:CorrelationDemo}. This is an important validation; the
uncertainties of individual resonances should not be affected by this
procedure, only their inter-dependence. Finally, correlated samples of
resonance strengths are calculated using
Eq.~(\ref{eq:res-strength-sample}). Each set of sampled strengths is
used to compute a sample reaction rate at temperature, T using
Eq.~(\ref{eq:reactionrate-narrowresonance}). The resulting rate
uncertainties are presented later in
Sec.~\ref{sec:results-discussion}.

\begin{figure}
  \centering
  \includegraphics[width=0.45\textwidth]{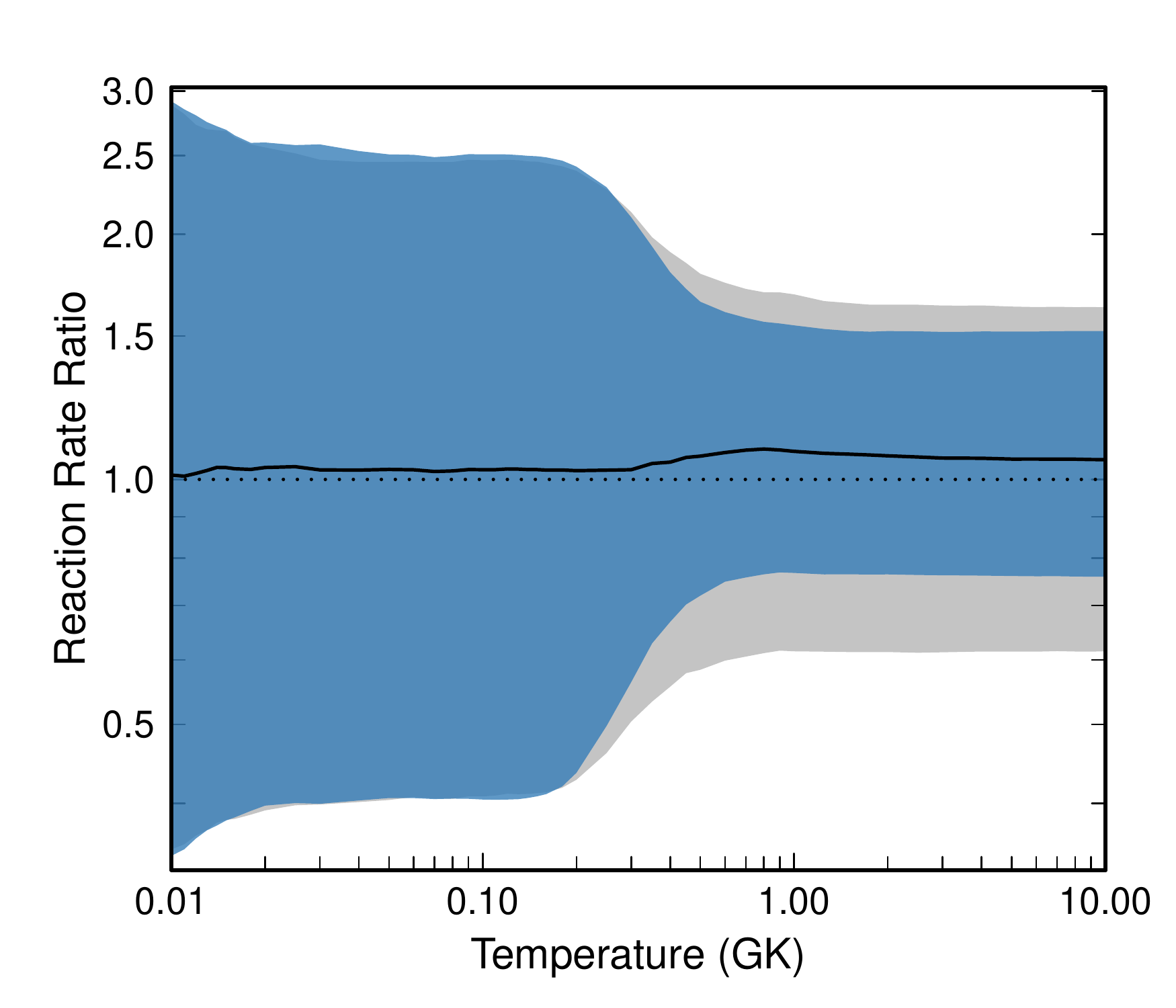}
  \caption{(color online) Reaction rate uncertainties as a function of
    temperature and normalized to the recommended rate. The blue
    shaded region shows the uncorrelated $1-\sigma$ uncertainties of
    the rate, while the grey region represents rate uncertainties once
    correlated uncertainties are taken into account. Clearly, at low
    temperatures where the rate is dominated by only one resonance,
    there is no difference in the reaction rate uncertainties when
    applying this procedure.}
  \label{fig:PlotCompare-Demo}
\end{figure}

The reaction rate uncertainties for the simple 3-resonance reaction
are shown in Fig.~\ref{fig:PlotCompare-Demo}. For clarity, the
uncertainties are normalized to the recommended rate. A value of 2.0,
for example, corresponds to a rate that is twice the recommended
rate. At low temperatures, the rate is dominated by resonance 1 and
the reaction rate probability distribution is largely unchanged
following the correlation procedure. At high temperatures above about
200 MK, correlation between resonances 2 and $r$ becomes
important. The grey uncertainty band representing correlated rates is
clearly wider than the blue, uncorrelated rate uncertainties. Even at
higher temperatures, though, the rate uncertainties don't increase
significantly, despite the strong correlation between resonances 2 and
$r$. This is a consequence of the relative strength of the two
resonances. The latter resonance dominates the rate at high
temperatures, contributing roughly 60\% with the other two resonances
only contributing 30\% (resonance 2) and 10\% (resonance
1). 

\subsection{\reaction{27}{Al}{p}{$\gamma$}{28}{Si}}

The \reaction{27}{Al}{p}{$\gamma$}{28}{Si} reaction is of
astrophysical importance, strongly influencing the rate of
\nuc{27}{Al} production in stellar environments by affecting the
leakage of material out of the AlMg cycle in massive
stars~\cite{PRA96,HAR00,POW98}. Its long recognized importance has
lead to numerous experimental resonance strength studies, most
recently those of \cite{POW98}, \cite{CHR99}, and
\cite{HAR00}. The former of these results has been used in
\cite{ILI10a} to normalize other experimental resonance strength
determinations, thus introducing correlation between these
parameters. Furthermore, experimental information for this reaction
reaches high enough energies that theoretical Hauser-Feshbach rates
are not required to supplement the experimental information at high
temperatures. Here, we use the same input parameters as those in
\cite{ILI10b}, so the reader is referred to that publication for
details.

\begin{figure}
  \centering
  \includegraphics[width=0.5\textwidth]{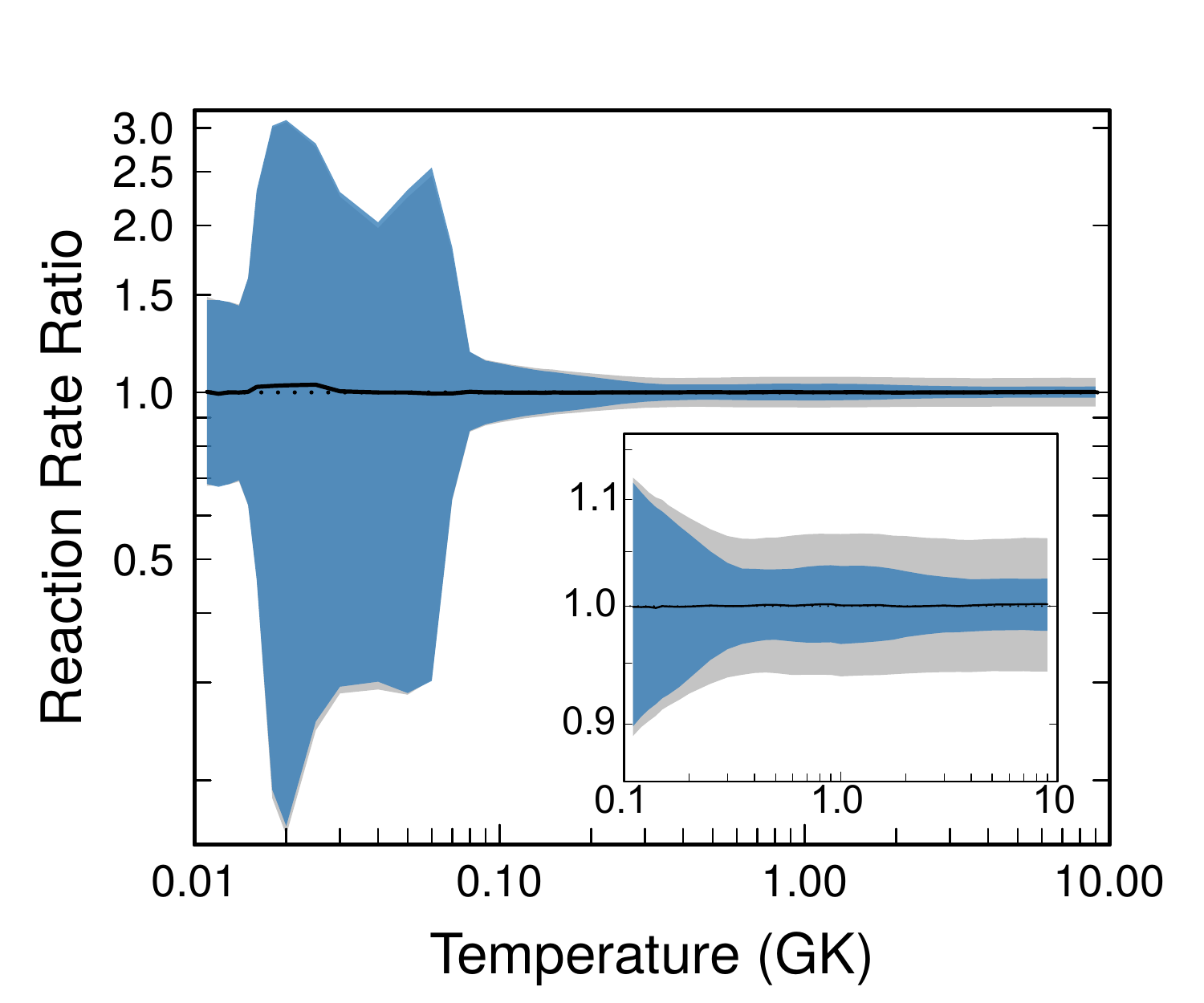}
  \caption{(color online) Reaction rate uncertainty comparison for the
    \reaction{27}{Al}{p}{$\gamma$}{28}{Si} reaction when including
    resonance strength correlations. Correlations only affect the
    reaction rate uncertainties at temperatures above 300 MK. This is
    because at lower temperatures, the reaction rate is dominated by
    only a few resonances at 72 keV, 84 keV, and 196 keV.}
  \label{fig:27Alpg-compare}
\end{figure}

Although the resonance density of the
\reaction{27}{Al}{p}{$\gamma$}{28}{Si} reaction is high, the
low-energy resonances that dominate the reaction rate below 100 MK are
poorly known and therefore are not strongly correlated with the
reference resonance at \Ercm{391.3} measured by
\cite{POW98}. Thus, the reaction rate uncertainties at
temperatures below 100 MK are largely unchanged when correlations are
taken into account. This is reflected in
Fig.~\ref{fig:27Alpg-compare}. However, at high temperatures, where
there are many contributing resonances that \textit{are} correlated
with the reference resonance, we see the reaction rate uncertainties
increase by up to a factor of 2.5. 

The \reaction{27}{Al}{p}{$\gamma$}{28}{Si} reaction primarily affects
leakage of material out of the Mg-Al cycle and strongly determines the
synthesis of \nuc{27}{Al} in massive
stars~\cite{PRA96,HAR00,POW98}. In these environments, temperatures
are restricted to T$<90$~MK. For the sake of the present study, we
consider a single-zone model based on the core hydrogen-burning
parameters shown in Cavallo \textit{et al.}~\cite{CAV98}. That is:
$T=50$~MK and $\rho=600$~g$/$cm$^3$. It is clear, from inspection of
Fig.~\ref{fig:27Alpg-compare}, that the rate uncertainties for this
reaction at 50~MK are unchanged when taking correlations into
account. Therefore, nucleosynthesis (namely the destruction of
\nuc{27}{Al} in the environment) in this case is unaffected by
correlation effects.

\subsection{\reaction{23}{Na}{p}{$\alpha$}{20}{Ne}}

The \reaction{23}{Na}{p}{$\alpha$}{20}{Ne} reaction is of key
importance in understanding the destruction of \nuc{23}{Na} needed to
explain abundance anomalies in globular clusters
\citep{RAF04,HAL04,CES13}. The destruction of sodium requires high
temperature hydrogen burning, such as the shell burning found in
massive AGB stars \citep{VEN01,DAN02,DEN03}, rotating massive AGB
stars \citep{DEC05}, rotating massive stars \citep{PRA06,DEC07}, or
massive binaries \citep{DEM09}.

Recent experimental results \citep{CES13} for
the competing \reaction{23}{Na}{p}{$\gamma$}{24}{Mg} reaction showed
that the main source of nucleosynthesis uncertainty for \nuc{23}{Na}
now comes from the (p,$\alpha$) reaction's cross section. We should
also note, here, that there is a mistake in the reaction rate input
for this reaction presented on page 281 of \cite{ILI10b}. The
resonance at \Ercm{-303.0} was found in \cite{HAL02} to have a
spectroscopic factor of $C^2S=0.02$. Combining that with the single
particle reduced width of $\theta_{\textrm{sp}}^2 = 0.6$ from
\cite{ILI97} yields a reduced width $\theta^2 = 0.012$, in
contradiction with the value found in \cite{ILI10b}. We will
therefore use this updated value in the calculations below. The
sub-threshold resonance at \Ercm{-303.0} now has a negligible
contribution to the reaction rate at all temperatures, which lowers
the rate by approximately 10\% at T$=0.01 - 0.06$~GK. All other inputs
are the same as those found in \cite{ILI10b}.

\begin{figure}
  \centering
  \includegraphics[width=0.45\textwidth]{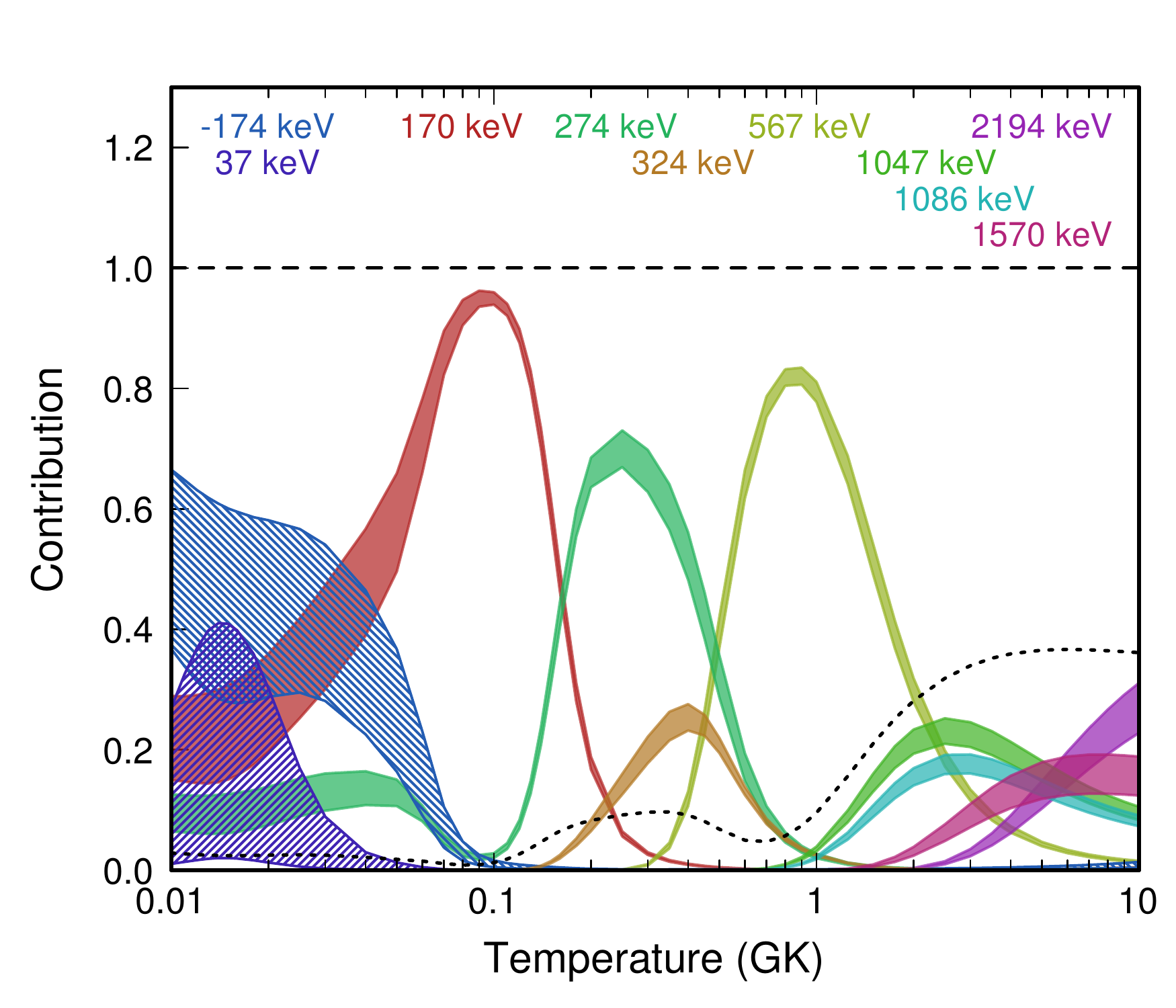}
  \caption{(color online) Fractional contributions of individual
    resonances and sub-threshold resonances to the
    \reaction{23}{Na}{p}{$\alpha$}{20}{Ne} reaction rate. The bands
    shown account for the uncertainties in each resonance's input
    parameters.}
  \label{fig:23Napa-contribution}  
\end{figure}

This reaction is rather more complex than the previous examples. It
consists of a number of directly measured resonances above \Ercm{217},
but also includes resonances and sub-threshold resonances whose widths
have been determined by some other means, namely particle transfer
measurements \citep{HAL04,FUC68}. We will assume here that the model
uncertainties in those measurements (\cite{ILI10b} uses a standard
value of 40\%) are correlated only by normalization to higher energy
resonances. Cross correlation between these transfer results would
arise from model effects, but these effects may affect states
differently, depending on their single-particle nature. The
sub-threshold resonances dominate the reaction rate at temperatures
below 40~MK while the two broad resonances at \Ercm{170} and
\Ercm{274} dominate the rate between 40~MK and about
500~MK. Fractional contributions to the rate are more readily
visualized in Fig.~\ref{fig:23Napa-contribution}. Care should be taken
in interpreting these contribution plots. For example, while the
\Ercm{170} resonance is rather uncertain (20\% uncertainty in the
proton width), its contribution to the total reaction rate at 100~MK
is well constrained at the 1\% level. The uncertainty in the
\textit{total} rate is therefore dominated by this single resonance at
100~MK (i.e. 20\% uncertainty). We expect the rate uncertainty from
correlations to vary significantly according to temperature in this
case.

\begin{figure}
  \centering
  \includegraphics[width=0.45\textwidth]{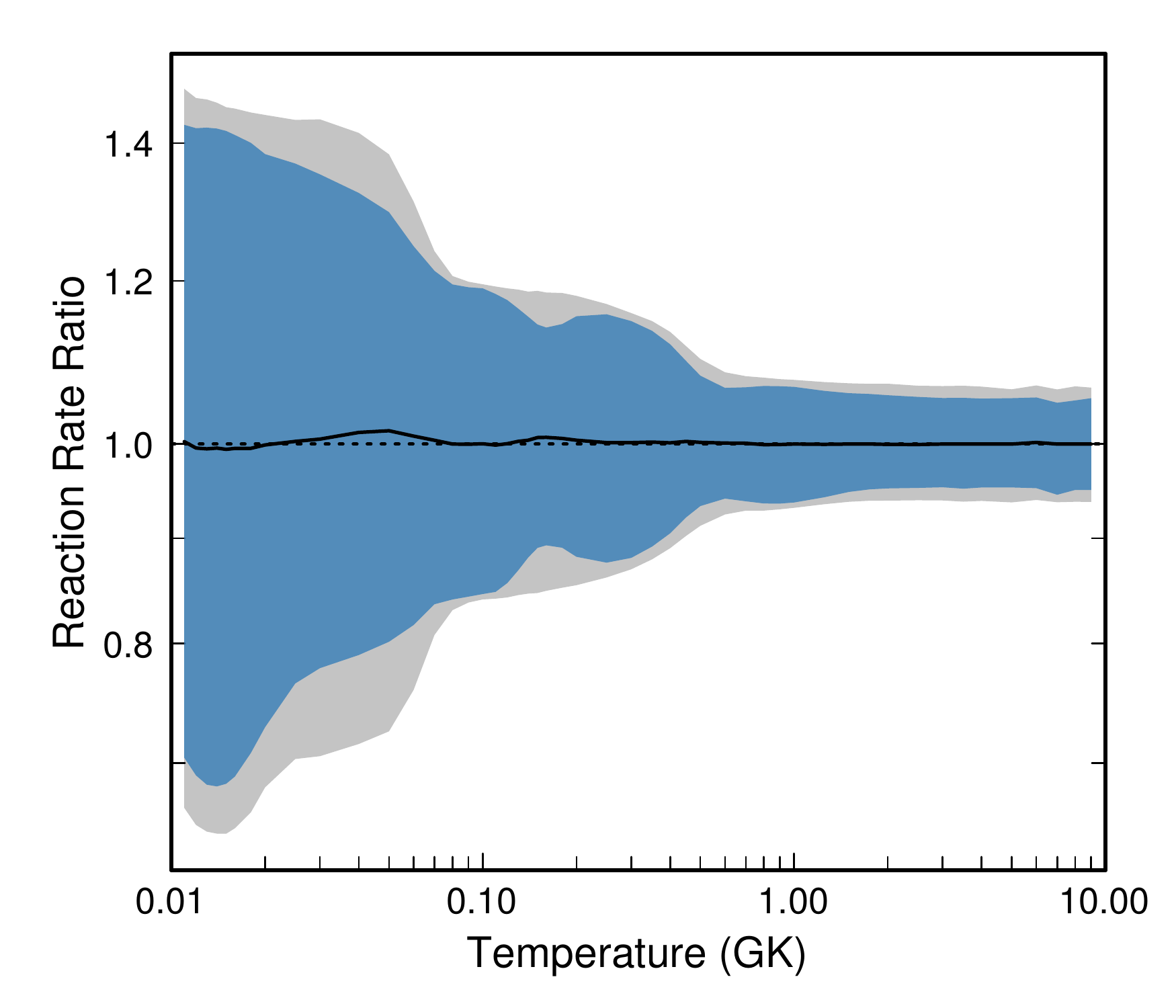}
  \caption{(color online) Reaction rate uncertainty comparison for the
    \reaction{23}{Na}{p}{$\alpha$}{20}{Ne} reaction. The blue and grey
    shaded regions correspond to the $1-\sigma$ reaction rate
    uncertainty bands for the uncorrelated and correlated Monte Carlo
    reaction rate calculations, respectively. }
  \label{fig:23Napa-compare}
\end{figure}

The rate of the \reaction{23}{Na}{p}{$\alpha$}{20}{Ne} reaction is
dominated by overlapping wide resonances at low temperatures. We
expect larger rate uncertainties if correlations are accounted for, as
is reflected in Fig.~\ref{fig:23Napa-compare}. When correlations are
taken into account at 30~MK, the rate uncertainties increase by
33\%. On the other hand, at 1~GK where only one resonance at
\Ercm{567} dominates, the rate uncertainties are largely
unchanged. These results have been matched to Hauser-Feshbach
theoretical rates at 3.5~GK using the procedure outlined in
\cite{NEW08}.

Accounting for correlations in reaction rates is crucial, especially
at temperatures where multiple resonances contribute similar fractions
of the reaction rate. This is the case at 150~MK where the rate is
dominated by two wide resonance at \Ercm{170} and \Ercm{274} (see
Fig.~\ref{fig:23Napa-contribution}). These resonances have comparable
uncertainties in their proton partial widths and are normalized to a
common reference resonance \citep{ZYS91}. If the resonances were
allowed to vary independently, their interplay would cause
\textit{smaller} uncertainties at 0.15~GK as shown in blue
(essentially, the Monte Carlo variations in each resonance partially
cancel each other out). Since these resonances are normalized to a
common reference, they should co-vary under Monte Carlo calculations
(i.e., as one increases, so should the other). By properly taking
correlations into account, this artefact is removed. The correlated
uncertainty band shown in grey in Fig.~\ref{fig:23Napa-compare}
indicates that the method presented here better accounts for the rate
uncertainty for multiple contributing resonances.


Whether correlation effects change nucleosynthesis predictions depends
strongly on the model in question. For example, shell hydrogen burning
in a 5 M$_{\odot}$ AGB star with metalicity $z = 10^{-3}$
\citep{VEN05} occurs at about 90-100~MK. The rate at this temperature
is dominated by the \Ercm{170} resonance, hence nucleosynthesis
uncertainties are largely unaffected. If another model operated at
higher temperatures where correlations do have an effect, we could
expect nucleosynthesis prediction uncertainties to increase by up to
30\%.

\subsection{\reaction{23}{Na}{$\alpha$}{p}{26}{Mg}}

The \reaction{23}{Na}{$\alpha$}{p}{26}{Mg} reaction was shown to be
important for \nuc{26}{Al} production during C/Ne convective shell
burning in massive stars \citep{ILI11}. The critical temperature
region identified by for \nuc{26}{Al} production in C/Ne convective
shell burning is 1.25 GK.  The reaction is an important proton source
necessary for the \reaction{25}{Mg}{p}{$\gamma$}{26}{Al} reaction to
proceed. It is \textit{also} important for the production of
\nuc{27}{Al}, whose abundance is critical in aluminium isotopic ratio
measurements.  

This reaction has been studied directly by \cite{KUP63} and
\cite{WHI74} and subsequently in inverse kinematics by
\cite{ALM14,TOM15,HOW15}, and \cite{AVI16}. The latter studies were
not performed with high enough resolution to distinguish individual
resonances, and will therefore be discarded for the sake of this
illustrative example. Note, though, that the latter studies supersede
the forward reaction studies as \cite{ILI11} outlined: target
deterioration effects were not properly taken into account in earlier
studies. Comparing the reaction rates determined from the normal and
inverse kinematics studies suggests that the target effects in
\cite{KUP63} and \cite{WHI74} amount to about 37\%.

The $\alpha$-particle and proton separation energies for this reaction
are 10.092~MeV and 8.271~MeV, respectively. The large Q-values mean
that it proceeds through higher lying states, so we expect a high
level density. Indeed, the average resonance spacing found by
\cite{KUP63} and \cite{WHI74} is about 30~keV. Here, we use the values
quoted in those studies with no correction or renormalization
applied. The studies are, however, self-normalized. In this case, we
expect many resonances to contribute to the reaction rate at any given
temperature, so correlations between their strengths will be crucial.

\begin{figure}
  \centering
  \includegraphics[width=0.45\textwidth]{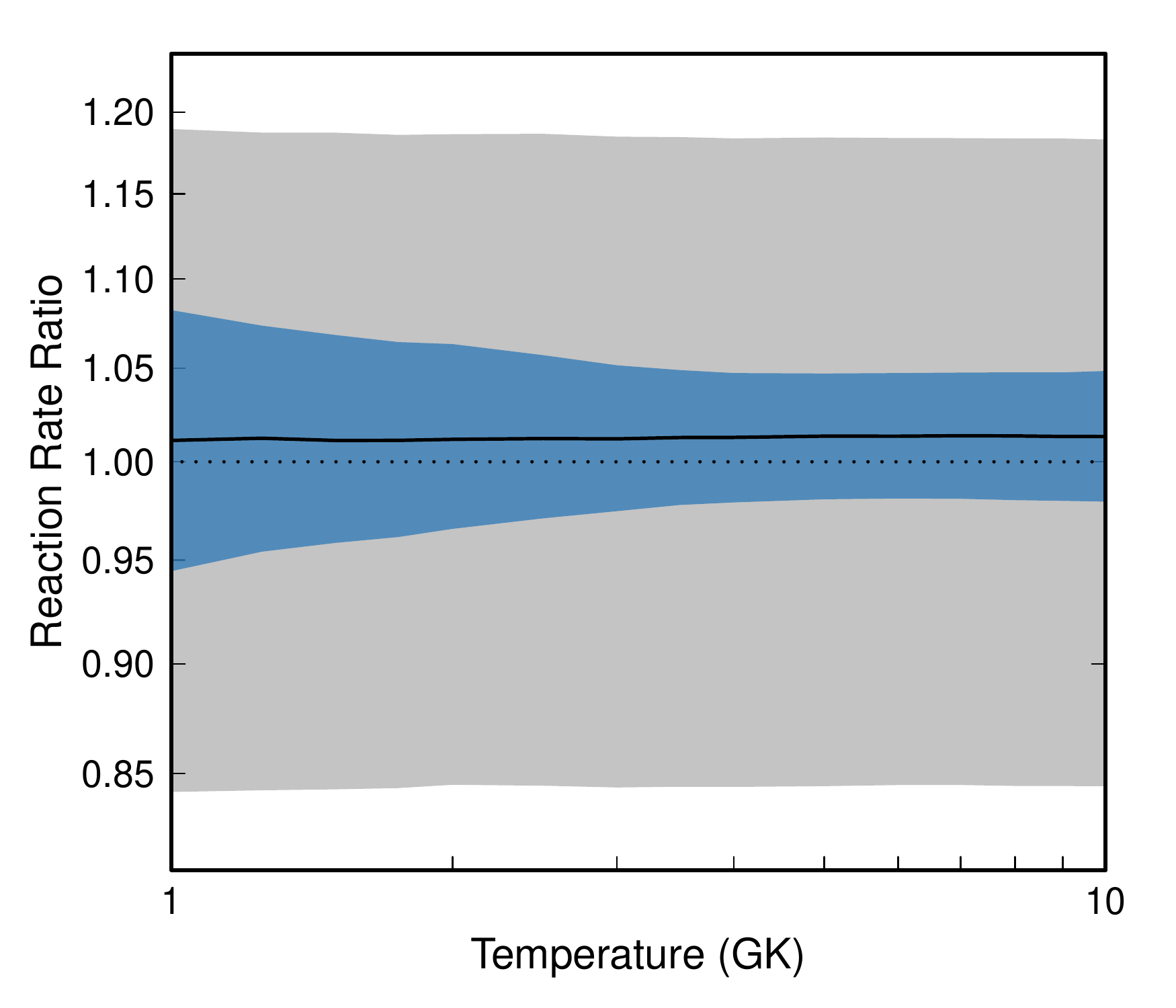}
  \caption{(color online) Reaction rate uncertainty comparison for the
    \reaction{23}{Na}{$\alpha$}{p}{26}{Mg} reaction. The blue and grey
    shaded regions correspond to the $1-\sigma$ reaction rate
    uncertainty bands for the uncorrelated and correlated Monte Carlo
    reaction rate calculations, respectively. }
  \label{fig:23Naap-compare}
\end{figure}

Rate uncertainties for the \reaction{23}{Na}{$\alpha$}{p}{26}{Mg}
reaction are shown in Fig.~\ref{fig:23Naap-compare} for temperatures
between 1 GK and 10 GK.  In this case, we see a very large effect of
correlations on the reaction rate uncertainties, where the correlated
uncertainties (shown in grey) are roughly a factor of 3 larger than
when the correlations are not properly taken into account (blue). For
higher temperatures, where even more resonances contribute to the
rate, the correlated uncertainty is up to a factor of 5 larger than
the uncorrelated case. This is because many resonances contribute to
the reaction rate owing to the high level density. Thus, their
experimental correlation is more critical.



\begin{figure}
  \centering
  \includegraphics[width=0.45\textwidth]{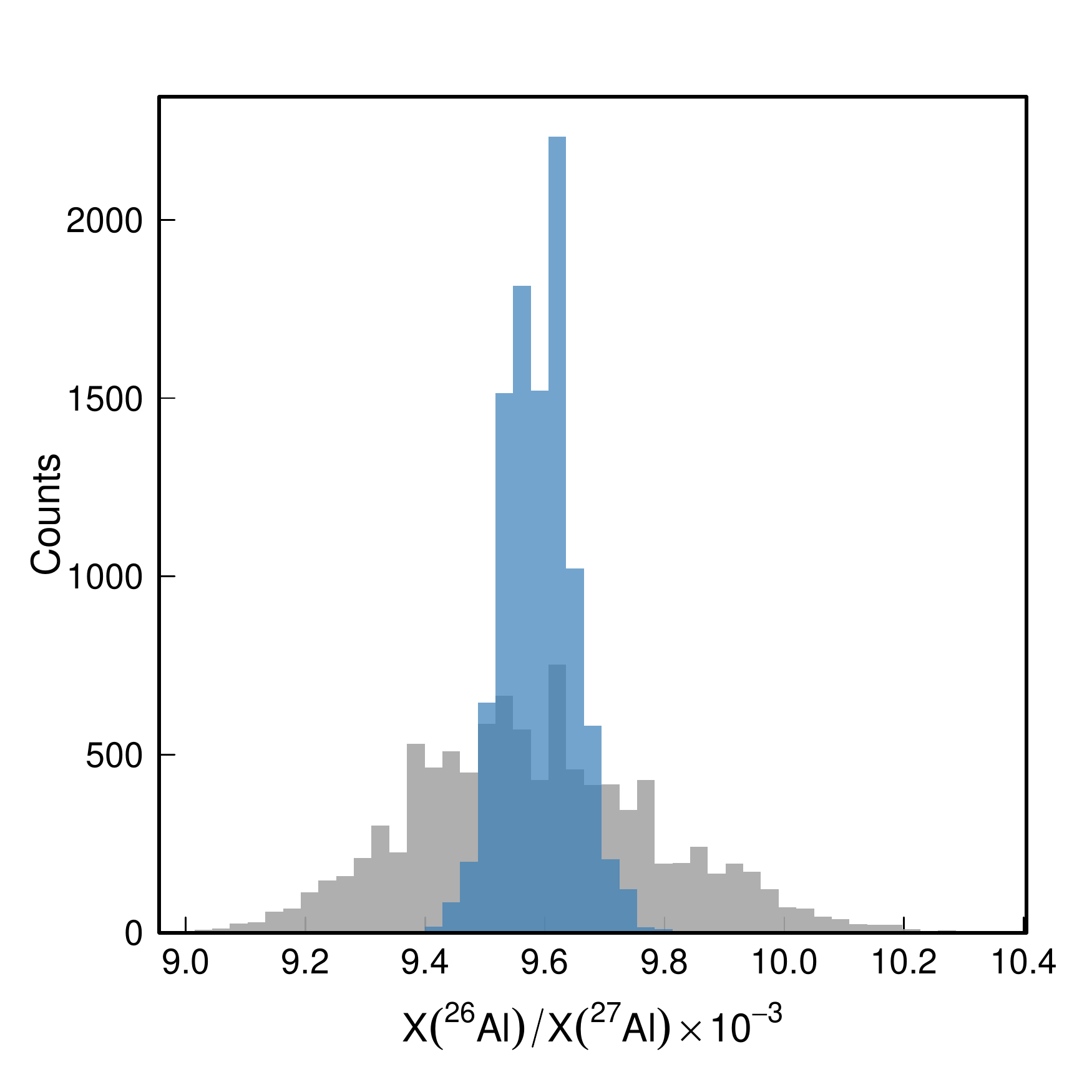}
  \caption{(color online) Final \nuc{26}{Al}/\nuc{27}{Al} abundance
    ratio for the C/Ne convective shell burning profile described in
    the text. Shown in blue and grey are the final Monte Carlo
    abundance ratio when considering uncorrelated and correlated
    reaction rates, respectively.}
  \label{fig:23Naap-26Al-over-27Al-comp}
\end{figure}

The temperature-density profile defined in \cite{ILI11} is
used to investigate the effect of correlated uncertainties on
aluminium production. In short, a single-zone profile is extracted
from the C/Ne convective shell of the 60 M$_{\odot}$ massive star
model in \cite{LIM06}. To account for mixing effects, the time
axis of this profile is then compressed by a factor of 60 to reproduce
the same nucleosynthesis pattern found in the full model. More details
can be found in \cite{ILI11}. The \nuc{26}{Al}/\nuc{27}{Al}
abundance ratio both before and after taking nuclear correlations into
effect is shown in Fig.~\ref{fig:23Naap-26Al-over-27Al-comp}.

The abundance ratio in this case is strongly affected by taking
correlations in experimental resonance strengths into account when
calculating reaction rates. Specifically, the uncertainty in the
abundance ratio of \nuc{26}{Al}/\nuc{27}{Al} increases by a factor of
3.5. Clearly, if precise measurements are to be compared with model
expectations, these correlation effects must accounted for.






\section{Summary}
\label{sec:summary}

Monte Carlo reaction rate calculations have introduced a new, powerful
tool for determining the uncertainties in rates arising from nuclear
physics uncertainties. These methods have allowed complex uncertainty
propagation to be included without the need for mathematical
simplifications, and opened the door for Monte Carlo nucleosynthesis
calculations. Before now, no attempt was made to take into account
correlations in those nuclear input parameters.

In this paper, the first study aimed at accounting for correlations in
resonance strengths and partial widths is presented. By assuming,
conservatively, that all resonance strengths or partial widths are
normalized to a reference resonance or common experimental systematic
error, we are able to compute correlated reaction rate uncertainties
with relatively little overhead. This method also establishes a
framework for computing correlation parameters in the absence of
published correlations. Indeed, in most cases, full covariance data
for a measurement is not published, so this conservative estimation
provides a safe strategy for flexible application.

Two main lessons can be learned from this investigation: (i) that at
low temperatures where most reactions are dominated by only a few
resonances, the Monte Carlo reaction rates computed using this method
are affected only slightly in comparison to the uncorrelated
assumption used previously, and (ii) where many resonances contribute
equally to the reaction rate, its uncertainties can increase
significantly.

In the future, it would be beneficial to consider the correlation of
resonance energies, also. This is particularly useful for radioactive
nuclei whose resonance energies are usually determined by subtracting
the reaction Q-value. However, it is more challenging in this case to
separate different experimental regimes. A few low-energy threshold
resonances with large uncertainties may be correlated with each other,
but not with a directly measured resonance at high energy.

Correlations between nuclear physics inputs helps us improve our
estimates of reaction rate uncertainties that are based on
experimental data. While this effect can be minor, it can have
dramatic effects on model predictions in certain cases. This study
represents a first step to more accurately represent rate
uncertainties.

\begin{acknowledgements} 
The author would like to thank Christian Iliadis, Art Champagne, and
Peter Mohr for their comments on the manuscript.
\end{acknowledgements}

\bibliographystyle{aa} 
\bibliography{CorrelatedUncertainties}

\end{document}